\documentclass[]{aa} 
\usepackage{graphicx}
\usepackage{aalongtable}
\usepackage{natbib}
\bibpunct{(}{)}{;}{a}{}{,}
\begin{document}
   \title{Results of the ROTOR-program}

   \subtitle{I. The long-term photometric variability of classical T
     Tauri stars\thanks{$UBVR$ photometric catalogues are available in
     electronic form at CDS Strasbourg via anonymous ftp to
     cdsarc.u-strasbg.fr.}}

   \author{K.N. Grankin \inst{1}, S.Yu. Melnikov \inst{1}, J. Bouvier \inst{2}, W. Herbst \inst{3}, \and V.S. Shevchenko
\inst{1, 4}}

   \institute{Astronomical Institute of the Uzbek Academy of Sciences, Tashkent, Uzbekistan, 700052
           \and
              Laboratoire d'Astrophysique, Observatoire de Grenoble, Universit$\acute{\rm e}$ Joseph Fourier, B.P. 53, F-38041 Grenoble
Cedex 9, France
           \and
              Astronomy Department, Wesleyan University, Middletown, CT 06459
           \and
              Deceased March 2000}

\date{Received .../ Accepted ...}

\abstract
{T Tauri stars exhibit variability on all timescales, whose
  origin is still debated.}
{We investigate the long term variability
  of CTTs over up to 20 years, characterize it from a set of
  statistical parameters and discuss its origin.}
{We present a unique,
  homogeneous database of photometric measurements for Classical T
  Tauri stars extending up to 20 years. The database contains more
  than 21\,000 $UBVR$ observations of 72 CTTs. All the data were
  collected within the framework of the ROTOR-program at Mount
  Maidanak Observatory (Uzbekistan) and together they constitute the
  longest homogeneous, accurate record of TTS variability ever
  assembled. We characterize the long term photometric variations of
  49 CTTs with sufficient data to allow a robust statistical
  analysis and propose an empirical classification scheme.}
{Several
  patterns of long term photometric variability are identified. The
  most common pattern, exhibited by a group of 15 stars which includes
  T Tau itself, consists of low level variability ($\Delta
  V\leq$0.4mag) with no significant changes occurring from season to
  season over many years. A related subgroup of 22 stars exhibits a
  similar stable long term variability pattern, though with larger
  amplitudes (up to $\Delta V\simeq$1.6 mag). Besides these
  representative groups, we identify three smaller groups of 3-5 stars
  each which have distinctive photometric properties.}
{The long term
  variability of most CTTs is fairly stable and merely reflects
  shorter term variability due to cold and hot surface spots. Only
  a small fraction of CTTs undergo significant brightness changes on
  the long term (months, years), which probably arise from slowly
  varying circumstellar extinction.}


\authorrunning{K.N. Grankin et al.}
\titlerunning{Results of the ROTOR-program. I. The long-term photometric variability of CTTs}

\maketitle
%

\section{Introduction}

T Tauri stars (TTs) were first identified as a class of irregular
variables by \citet{joy42, joy45}. The properties of these low mass
pre-main sequence stars were reviewed by, e.g., \citet{menard99} who
emphasized two subgroups, the so-called ``classical'' T Tauri stars
(CTTs) still actively accreting from their circumstellar disks and
the ``weak-line'' T Tauri stars (WTTSs) no longer surrounded by a
circumstellar disk. As discussed by \citet{herbst94}, three types of
day to week timescale variability can be identified in TTs. Type I
variability, most often seen in WTTSs, is characterized by a low level
periodic modulation of the stellar flux (a few 0.1 mag) and results
from the rotation of a cool spotted photosphere. Type II variations
have larger photometric amplitudes (up to 2 or 3 mag in extreme
cases), are most often irregular but sometimes periodic, and are
associated with short-lived accretion-related hot spots at the stellar
surface of CTTs. Finally, the more rarely observed Type III
variations are characterized by luminosity dips lasting from a few
days up to several months, which presumably result from circumstellar
dust obscuration, and are associated with the so-called UXOrs.

The ROTOR program (Research Of Traces Of Rotation) is dedicated to the
study of the photometric variability of pre-main-sequence (PMS) and
main-sequence (MS) objects. The goal of the program is to obtain a
long term homogeneous database of $UBVR$ observations of classical T
Tauri stars (CTTs), weak-line T Tauri stars (WTTs), post--T Tauri
stars (PTTs), Herbig AeBe stars (HAeBe), FU Orionis objects, and more
evolved stars such as RS CVn, BY Dra, and FK Com stars. A detailed
description of the program was given by \citet{shev89}. More than
100\,000 observations of about 370 variable stars have been obtained
since 1983. Our list contains 72 CTTs, 45 WTTs, 54 PTTs, 88 HAeBe, and
11 FK Com stars. Observations were secured over more than 20 years by
members of the Tashkent Astronomical Institute, including S.Yu.
Melnikov (24 \% of all observations), K.N. Grankin (23 \%), M.
Ibrahimov (14 \%), S.D. Yakubov (10 \%), O.V. Ezhkova (10 \%), and
V. Kondratiev (7 \%). Preliminary results were published by
\citet{shev93a, shev93b}, \citealt{gran95, shev98}, \citet{gran97,
gran98, gran99}, \citet{herbst99}.

In this paper, we analyze the unique and extensive Mt. Maidanak
database to investigate CTTs variability over a timescale of several
years and, in some cases, several decades. In Section 2, we
describe the stellar sample and the observations carried out for
the last 20 years at Mt. Maidanak. In Section 3, we define the
statistical parameters used to characterize the long term
variability of CTTs. In Section 4, we provide the results of the
statistical analysis of CTTs light curves and offer in Section 5
an empirical classification of their long term photometric
patterns. In Section 6, we attempt to relate the various types of
photometric patterns seen to the underlying causes of variability.

\section{Star sample and observations}

All $UBVR$ data were obtained at Mount Maidanak Observatory
(longitude: E$4^{\rm h}27^{\rm m}35^{\rm s}$; latitude:
$+38\degr41\arcmin$; altitude: 2540 m) in Uzbekistan. Results of
astroclimate studies at Maidanak observatory have been
summarized by \citet{ehgamberd00}, who show that it is among the
best observatories in the world, in many respects. We targeted
72 CTTs for long-term monitoring from the list of
\citet{herbig88}. More than 21\,000 $UBVR$ magnitudes were
collected for these objects from 1983 up to 2003, although the
number of U magnitudes is relatively small compared to the other
colors due to the faintness of the stars at those wavelengths.
All the observations were obtained at two telescopes (0.6 and
0.48 m reflectors) using a single-channel pulse-counting
photometer with photomultiplier tubes. As a rule, each program
star was measured once per night at minimal airmass. We secured
up to 120 $(U)BVR$ measurements for each sample target during
each observational season lasting several months. The number of
measurements depends on the object's visibility from Mt.
Maidanak. The largest number of measurements was thus obtained
for CTTs in Cygnus, Cassiopea, and Taurus, and fewer in Ophiuchus
and Orion. Magnitudes are provided in the Johnson $UBVR$ system.
The rms error of a single measurement in the instrumental system
for a star brighter than $12^m$ in $V$ is about $0^m.01$ in
$BVR$ and $0^m.05$ in $U$.

Observations were carried out either differentially using a nearby
reference star \citep[see][]{straizis77} or directly by estimating the
nightly extinction \citep{nikonov76}. In the latter case, several
reference stars were observed every night to derive the extinction
coefficients in each filter. Selected standard stars
\citep{landolt83,landolt92} were observed and used to calibrate
instrumental magnitudes on the Cousins system. We then transformed the
magnitudes to the Johnson $UBVR$ system using the relationship from
\citet{landolt83}~: (V-R)$_c$ = -0.0320 +
0.71652$\times$(V-R)$_J$. The formal accuracy of this reduction step
is $0^m.01$. All observed local times have been converted to
heliocentric julian days (JDH). A detailed description of the
equipment, observation techniques, and data processing is given in
\citet{shev89}.

The results of the Maidanak program of homogeneous long-term
photometry of CTTs are summarized in Table~1. Columns are: star's
name, star's number in \citet{herbig88}'s catalogue, spectral type,
time span of observations in JD, number of seasons observed,
photometric range in the $V$ band, number of observations in the $V$
band, and average values of $B-V$ and $V-R$ colors. The complete
Maidanak database for CTTs is available in electronic form at CDS,
Strasbourg, via anonymous ftp to cdsarc.u-strasbg.fr.

\section{Statistical parameters}

We define here the statistical parameters which allow us to
characterize the long term photometric variability of CTTs. To
increase the reliability of the statistical analysis we consider only
observing seasons having at least 12 measurements and further exclude
those stars which have less than 5 seasons of observations. As a
result we retain 49 CTTs for the analysis. We also performed a more
stringent analysis by retaining only 39 stars having more than 20
measurements per season and more than 5 seasons of observations. The
results of the 2 analyzes are not significantly different from one
another. We thus conclude that our selection criteria for the first
sample of 49 stars provide robust statistical parameters (mean,
standard deviations, minimum and maximum levels, etc.).

For each object, we compute the mean light level (${V_{m}}^{(i)}$)
in season~($i$) and the corresponding photometric range ($\Delta
V^{(i)}$) for that season. Then we average the  (${V_{m}}^{(i)}$)
over all seasons to yield ($\overline{V_m}$) and also compute its
standard deviation ($\sigma_{V_m}$) based on the scatter of the
seasonal means. Similarly, we compute the average $V$ band
photometric amplitude over all seasons ($\overline{\Delta
V}=\frac{\sum \Delta V^{(i)}}{N_s}$, where $N_s$ is the number of
observational seasons) and its standard deviation ($\sigma_{\Delta
V}$). We also use the ratio $\frac{\sigma_{\Delta
V}}{\overline{\Delta V}}$.

In addition to statistical parameters, we investigate the color
behavior of CTTs. Previous studies of the short term variability of
CTTs have shown that most stars tend to redden in color indices (such
as $V-R$ and $B-V$) when fading \citep[e.g.][]{herbst94}. This
behavior can result from surface spots, either cold or hot
\citep[cf.][]{bouvier95}, or from variable extinction by circumstellar
dust. We characterize the color behavior of CTTs by computing the
color slopes $\frac{\Delta(B-V)}{\Delta V}$ and
$\frac{\Delta(V-R)}{\Delta V}$ from least-square fits, and the
correlation coefficients $\rho_{B-V}$ and $\rho_{V-R}$. We used the
full set of data pertaining to each object in order to compute these
parameters since there are usually too few measurements per season to
obtain accurate seasonal color slopes. For most stars, a strong
correlation is found between color and brightness variations with
little evidence for significant variations from season to season (cf.,
e.g., Fig.~8).

The resulting values of statistical and color parameters for 49 CTTs
are listed in Table~2. The columns in Table 2 provide the star's name,
the number of seasons $N_s^*$ with at least 12 measurements, the mean
brightness level ($\overline{V_m}$) and its seasonal standard
deviation ($\sigma_{V_m}$), the average photometric range
($\overline{\Delta V}$, averaged over all seasons) and its seasonal
standard deviation ($\sigma_{\Delta V}$), the fractional variation of
photometric amplitude as measured by the ratio $\frac{\sigma_{\Delta
V}}{\overline{\Delta V}}$, and the color slopes $\frac{\Delta
(B-V)}{\Delta V}$ and $\frac{\Delta (V-R)}{\Delta V}$ together with
their correlation coefficients $\rho_{B-V}$ and $\rho_{V-R}$. Two
additional parameters discussed below (cf. Sect.4) are also listed,
which characterize the relative seasonal variability of the maximum
and minimum light levels ($C2 = \sigma_{V_{min}}/\sigma_{V_{max}}$)
and the preferred brightness state of the object on the long term ($C1
= <\frac{V_{med} - V_{min}}{\Delta V}>$ where $V_{med}$ is the median
light level of the object in a given season and the bracket indicates
an average over all seasons). For instance, a low C1 value indicates
an object which spends more time close to its maximum brightness
level.

\begin{figure}
\sidecaption
\centering
\includegraphics[width=9cm]{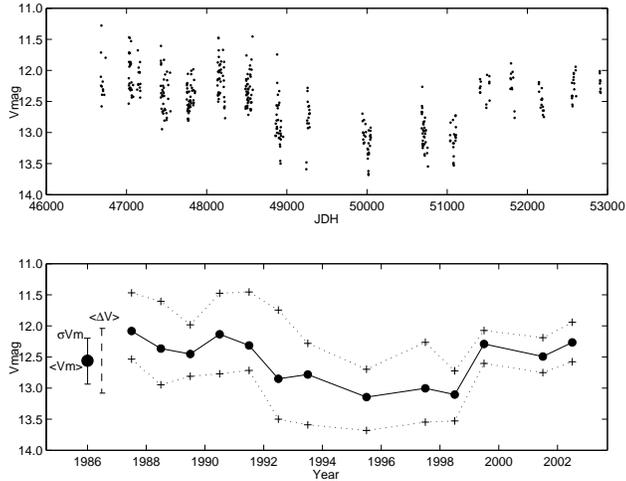}
\caption{The long term V-band light curve of the CTTs UY Aur. Top panel:
  photometric measurements; lower panel: statistical analysis (see text
  for details).} \label{Fig1}
\end{figure}

In order to illustrate the meaning of these parameters, an example of
the statistical analysis is shown in Figure~1 for \object{UY Aur}.
The long-term V-band light curve obtained at Maidanak Observatory is
shown in the upper panel. In the lower panel, some of the statistical
parameters are illustrated. The mean light level for each season of
observations (${V_{m}}^{(i)}$) is marked by a filled circle and the
minimum and maximum brightness by crosses.  Statistical parameters are
not computed for two seasons (2001 and 2004) which contain less than
12 measurements. The solid line joins values of average brightness and
demonstrates significant changes on years timescale. The photometric
amplitude is also seen to vary on the long term.

Figures~2a and 2b display the light curves of the other 48 CTTs on
which we applied the same statistical analysis. It is immediately
obvious that different photometric patterns occur on the long term. We
discuss these patterns with the help of the above defined statistical
parameters in the next section.

\begin{figure*}
\centering
\includegraphics{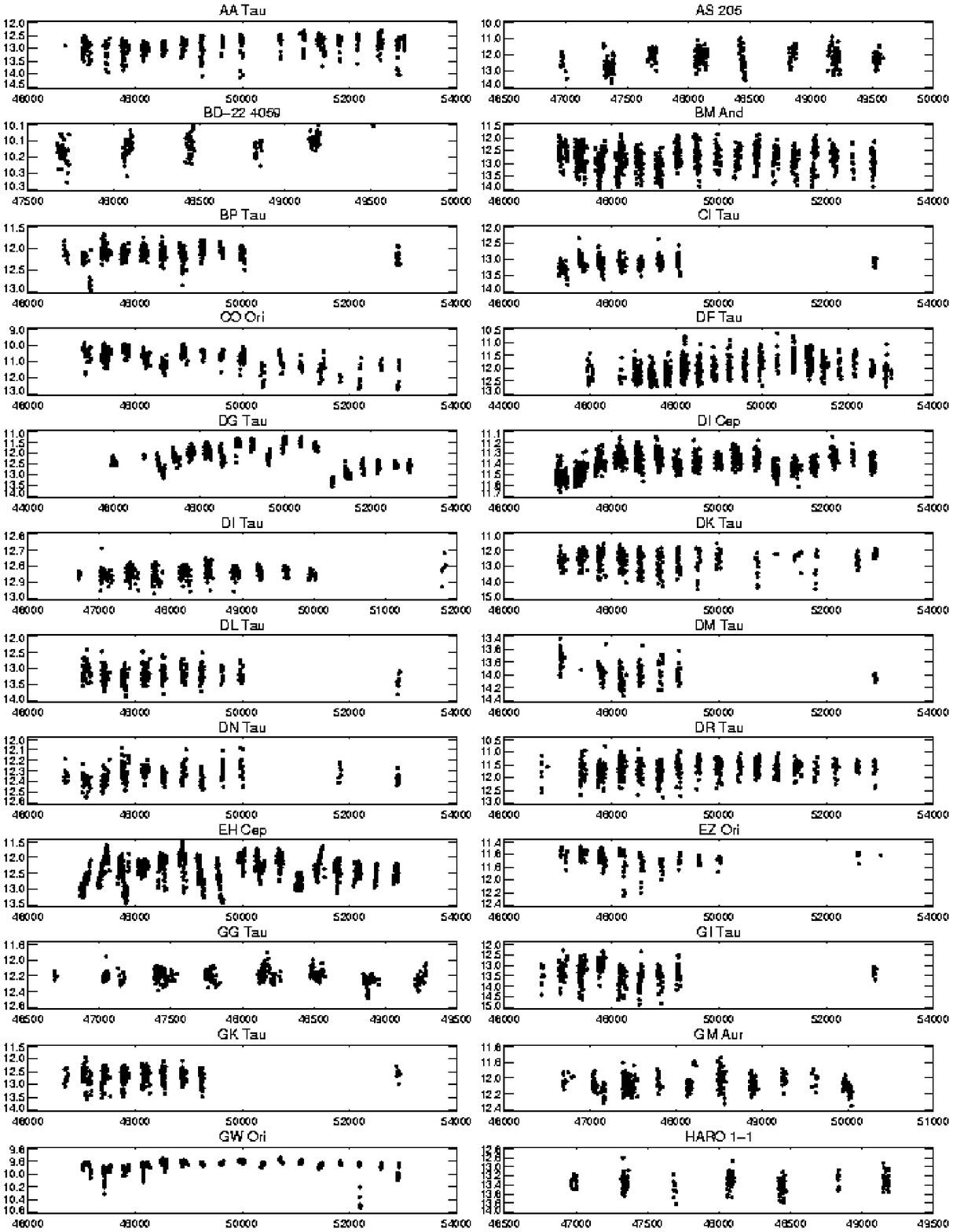}
Fig.2a Long term V-band light curves of 48 CTTs observed at Mt. Maidanak. \vspace{13cm}
\end{figure*}

\begin{figure*}
\centering
\includegraphics{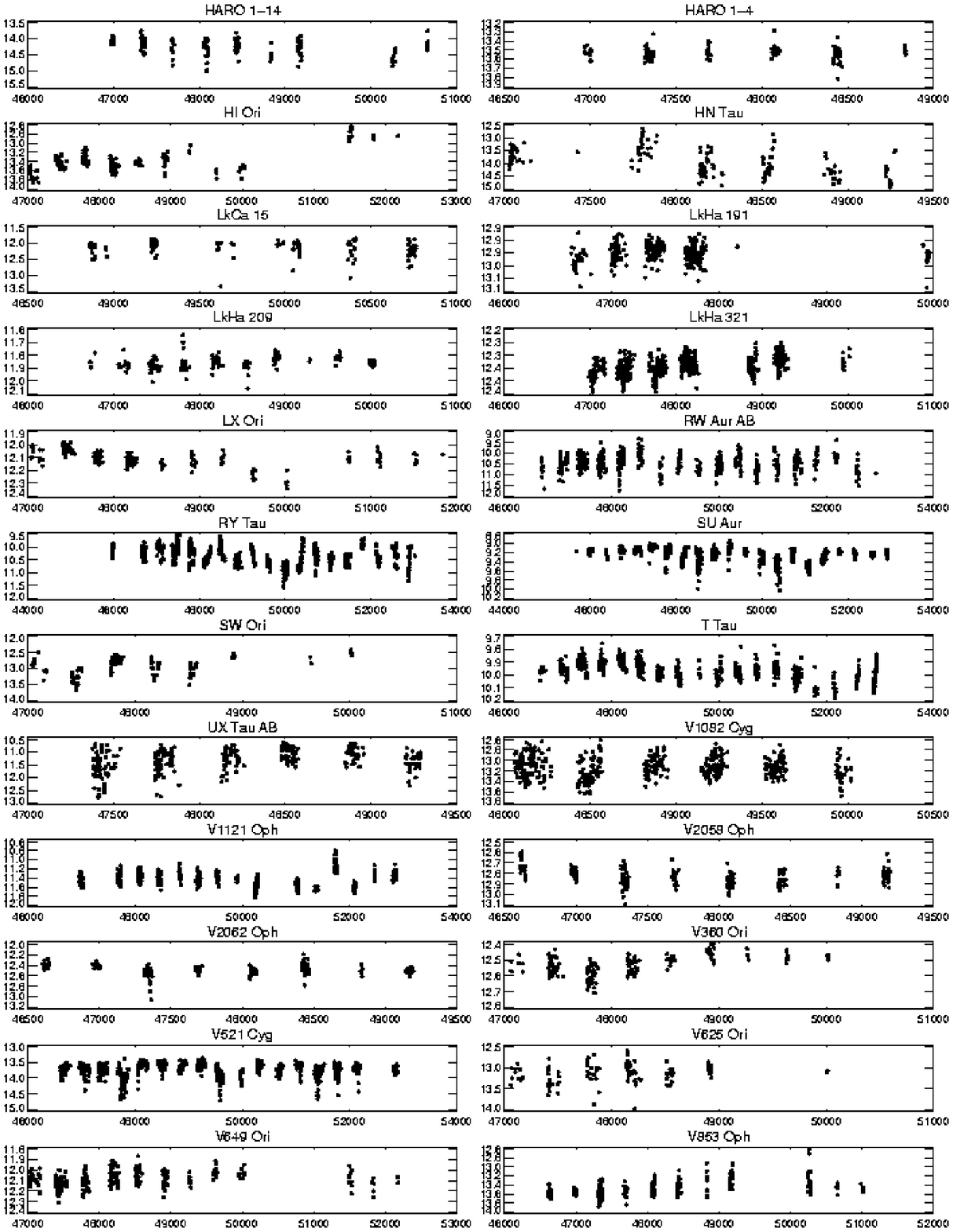}
Fig.2a (ct'd).
\end{figure*}

\begin{figure*}
\centering
\includegraphics{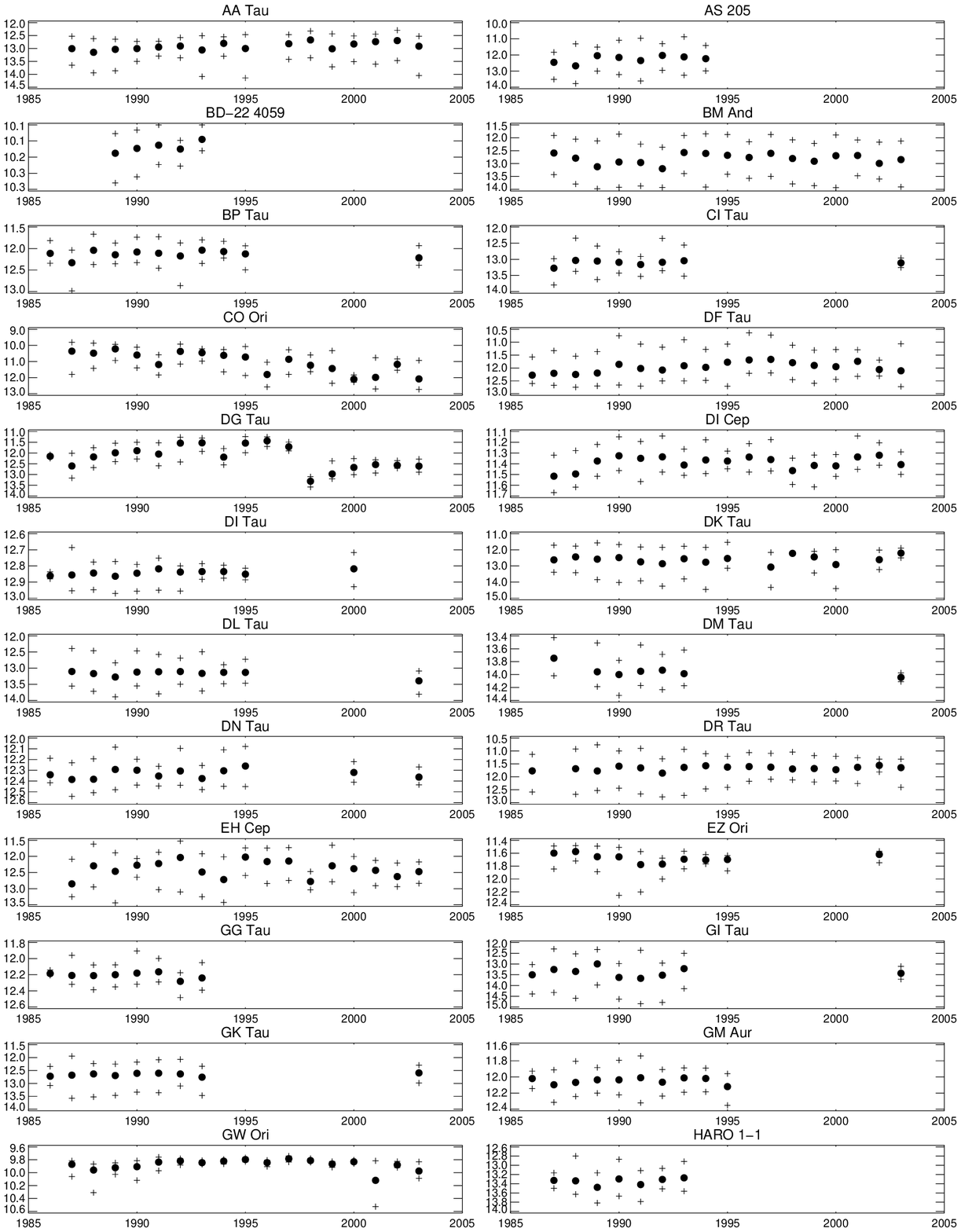}
Fig.2b Statistical representation of the 48 CTTs light curves shown in
Fig.2a
\end{figure*}

\begin{figure*}
\centering
\includegraphics{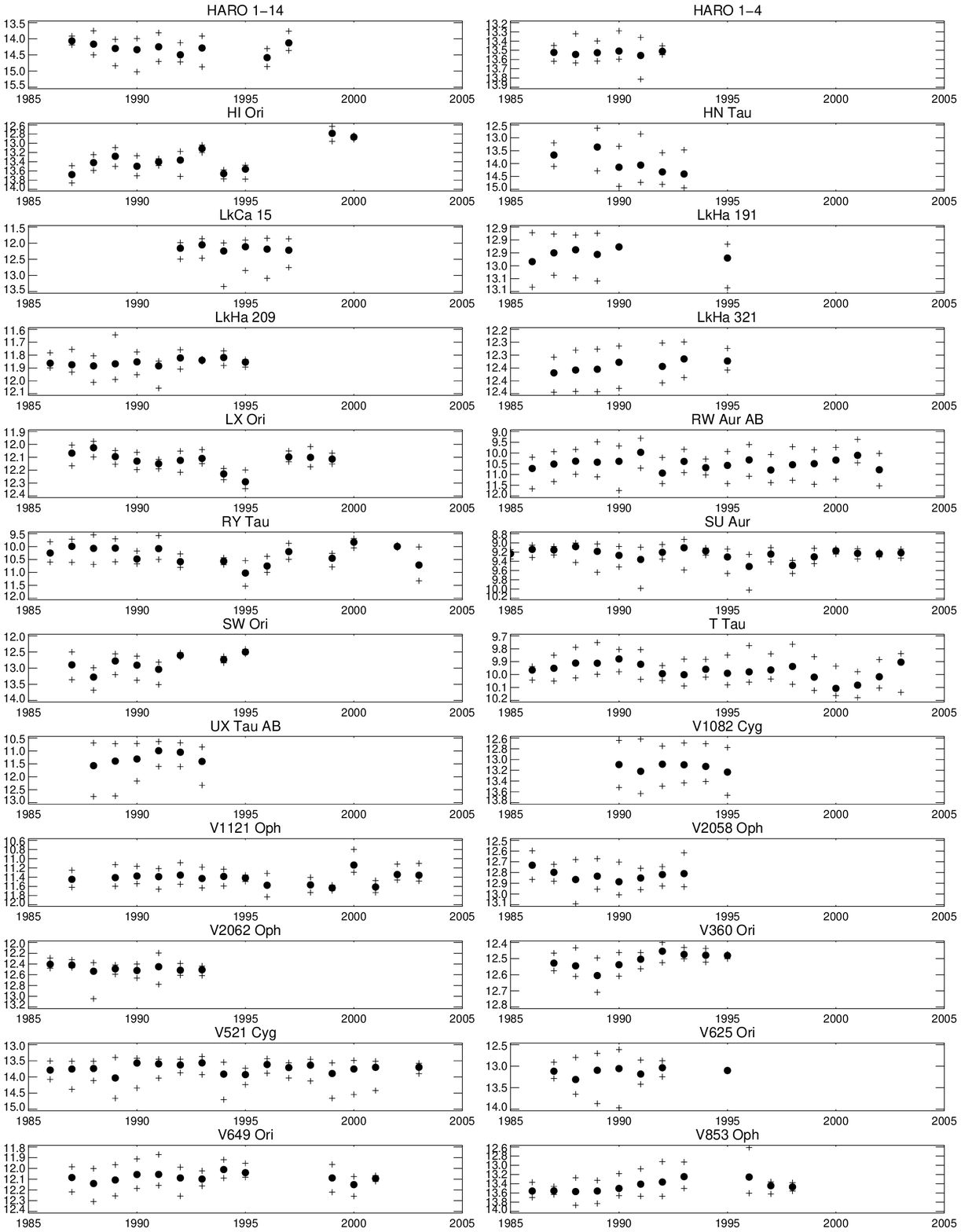}
Fig.2b (ct'd).
\end{figure*}

\begin{figure*}
\begin{minipage}[t]{8.7cm}
\includegraphics[width=8.5cm]{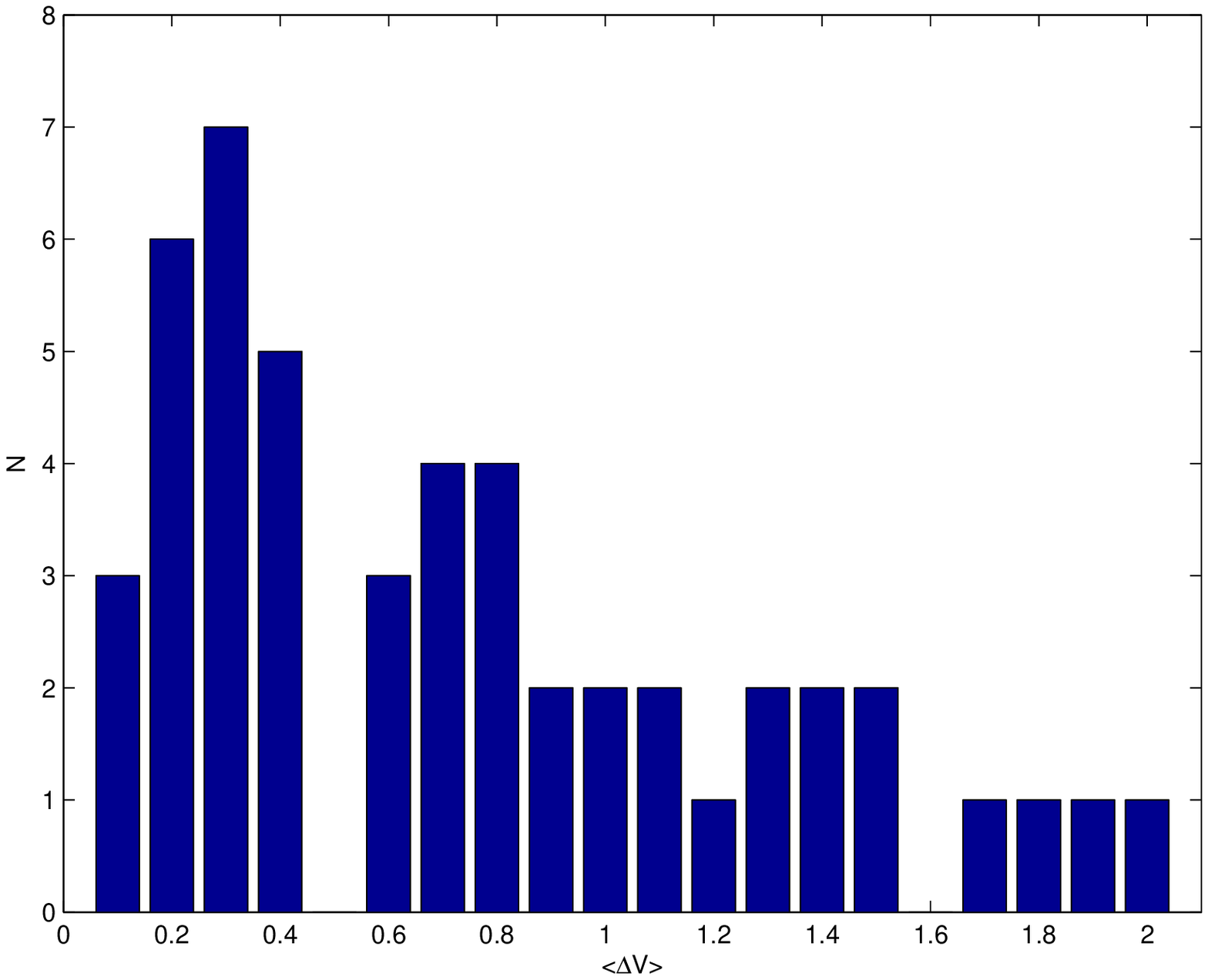} Fig.3. Average photometric
  amplitude in the $V$-band for the 49 CTTs of our sample
  ($\overline{\Delta V}$).
\end{minipage}
\hfill
\begin{minipage}[t]{8.7cm}
\includegraphics[width=8.5cm]{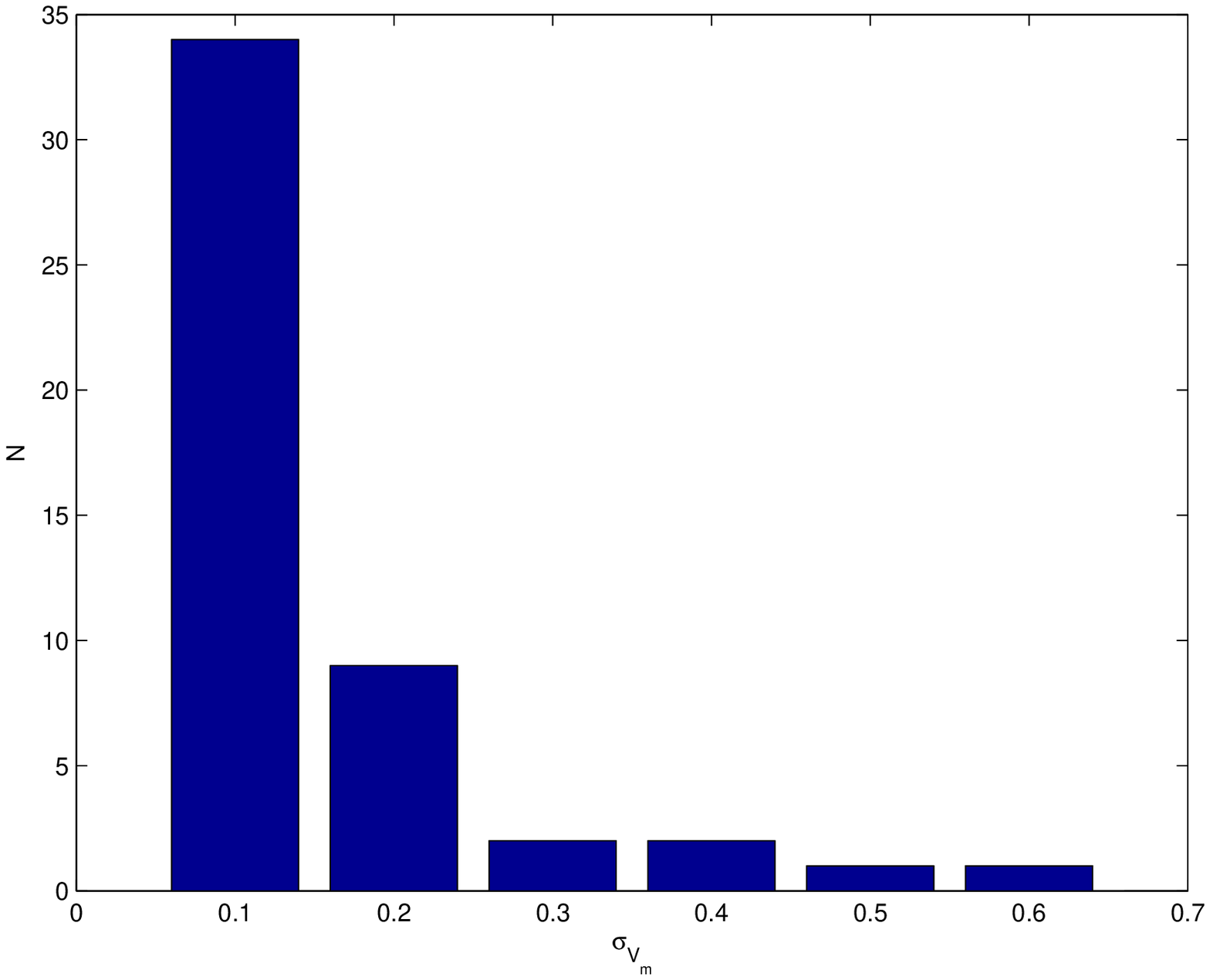} Fig.4. Standard deviation of the
  seasonal mean brightness level ($\sigma_{V_m}$).
\end{minipage}
\end{figure*}

\begin{figure*}
\begin{minipage}[t]{8.7cm}
 \includegraphics[width=8.5cm]{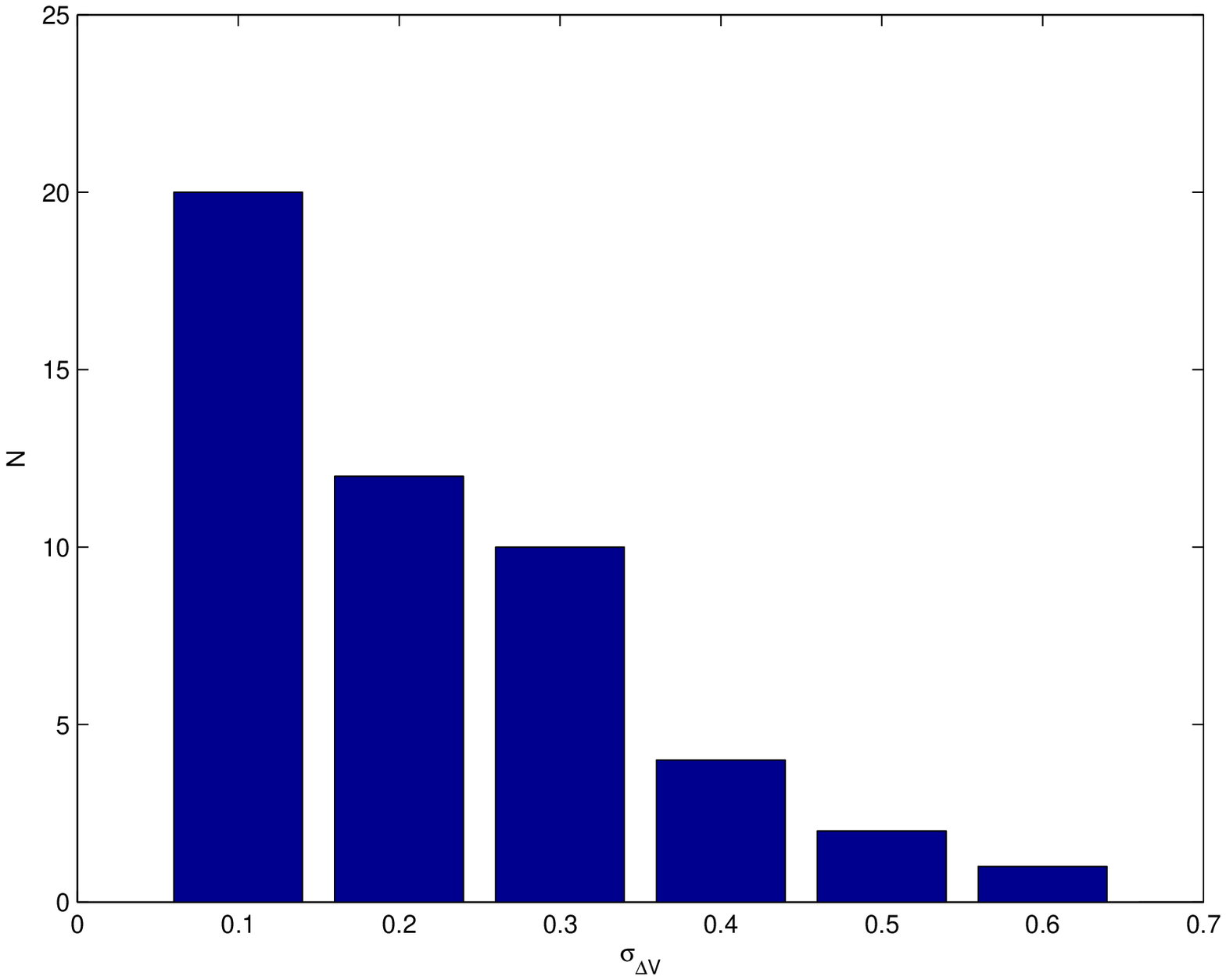} Fig.5. Standard deviation of
  the seasonal photometric amplitude ($\sigma_{\Delta V}$).
\end{minipage}
\hfill
\begin{minipage}[t]{8.7cm}
 \includegraphics[width=8.5cm]{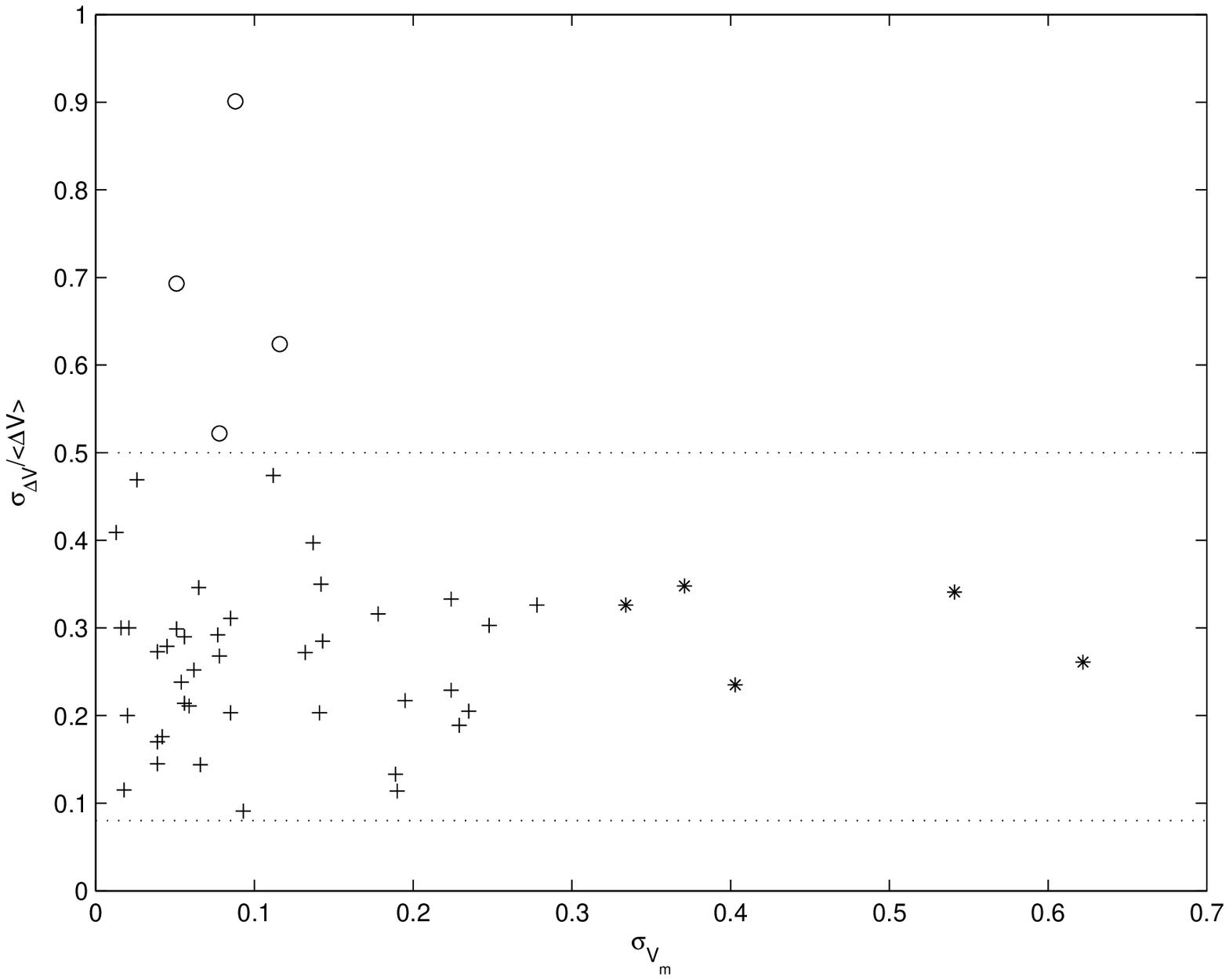} Fig.6. The relative variation
  of amplitude vs. the variation of the mean brightness level.  Four objects with the
  largest relative variations of amplitudes are shown by empty
  circles. Five stars
  with the largest changes in mean light level are denoted by asterisks.
\end{minipage}
\end{figure*}

\section{Results}

In this section, we present the results of the statistical analysis of
the long term light curves.  The histogram of the average V-band
amplitude of variability for the CTTs in our sample is shown in
Figure~3. The distribution is rather large, with $\overline{\Delta V}$
ranging from $0^m.1$ up to $2^m$.

Nearly half of the CTTs (21) exhibit small ranges of variability on
the long term, from $0^m.1$ to $0^m.4$. A roughly equal number (24)
display larger ranges, from $0^m.6$ to $1^m.5$ in the V-band.  Only a
few CTTs (4) exhibit average V-band ranges larger than $1^m.5$ and up
to $2^m$ over the years. These are \object{BM And}, \object{GI Tau},
\object{AS 205}, and \object{DK Tau}.

A histogram showing the frequency distribution of the measured
variations of mean light level from season to season of the 49 CTTs in
our statistical sample is displayed in Figure~4.  Specifically, this
variation is measured by the standard deviation of the seasonal mean
light levels ($\sigma_{V_m}$). The histogram is strongly peaked, with
the large majority of CTTs demonstrating little variation in their
mean light level over the years ($\sigma_{V_m}\simeq
0^m.1-0^m.2$). However, a few CTTs (6) are characterized by
significant changes in average brightness, with $\sigma_{V_m}$ ranging
from $0^m.25$ to $0^m.65$ (cf. Fig.~6). These are CO Ori, DG Tau, HN
Tau, UY Aur, RY Tau, and HI Ori. Figures~1 and 2 show that the mean
light level can vary smoothly on a timescale of several years (e.g. UY
Aur) or, on the contrary, can change quite abruptly from one season to
the next (e.g. DG Tau).

Similarily, the variations of the V-band range from season to season
are small for most CTTs, amounting to only a few tenths of a magnitude
at most, as shown by the histogram of $\sigma_{\Delta V}$ in
Figure~5. We caution that in seasons where the number of data is not
large or in the case of stars where the variations are rapid and
short-lived, we may be underestimating the true photometric
range. Nonetheless, it appears that this histogram is not as peaked
towards small values as is the histogram of $\sigma_{V_m}$ (Fig.4) but
displays a more continuous distribution. We find that the relative
changes in amplitude ($\frac{\sigma_{\Delta V}}{\overline{\Delta V}}$)
are usually constrained to lie between 0.1 and 0.5, except for 4 CTTs
which do exhibit larger seasonal variations up to 0.9 (EZ Ori, SU Aur,
V2062 Oph, and GW Ori, see Fig.6). In other words, CTTs with the
largest photometric amplitudes are also those whose amplitude varies
the most on a timescale of years.

That most CTTs exhibit relatively similar variations in mean
brightness level and fractional amplitudes is best seen in Figure~6
where the 2 quantities are plotted against each other. The vast
majority of CTTs in our sample lie in a small region of the diagram
delimited by $\sigma_{V_m}\leq 0^m.3$ and $\frac{\sigma_{\Delta
V}}{\overline{\Delta V}}\leq 0.5$. The few deviant CTTs, already
mentionned above, are easily identified in this plot.

\begin{figure*}
\begin{minipage}[t]{8.7cm}
 \includegraphics[width=8.5cm]{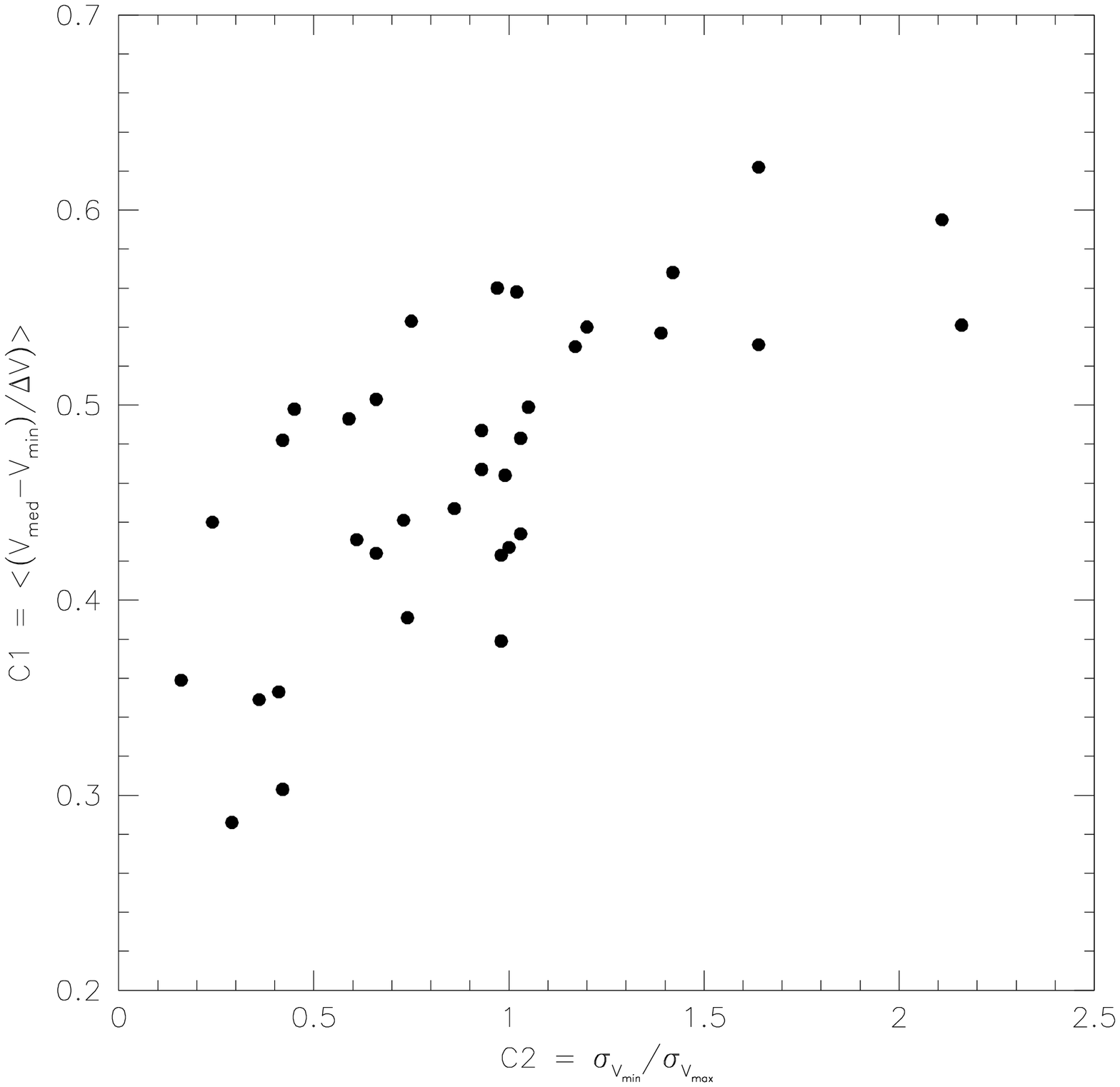} Fig.7. The star's preferred
  brightness state (as measured by C1=${<\frac{V_{med} - V_{min}}{\Delta
      V}>}$) is plotted against the relative variations of its extreme
  brightness states (as measured by
  C2=$\sigma_{V_{min}}/\sigma_{V_{max}}$). See text.
\end{minipage}
\hfill
\begin{minipage}[t]{8.7cm}
\includegraphics[width=8.5cm]{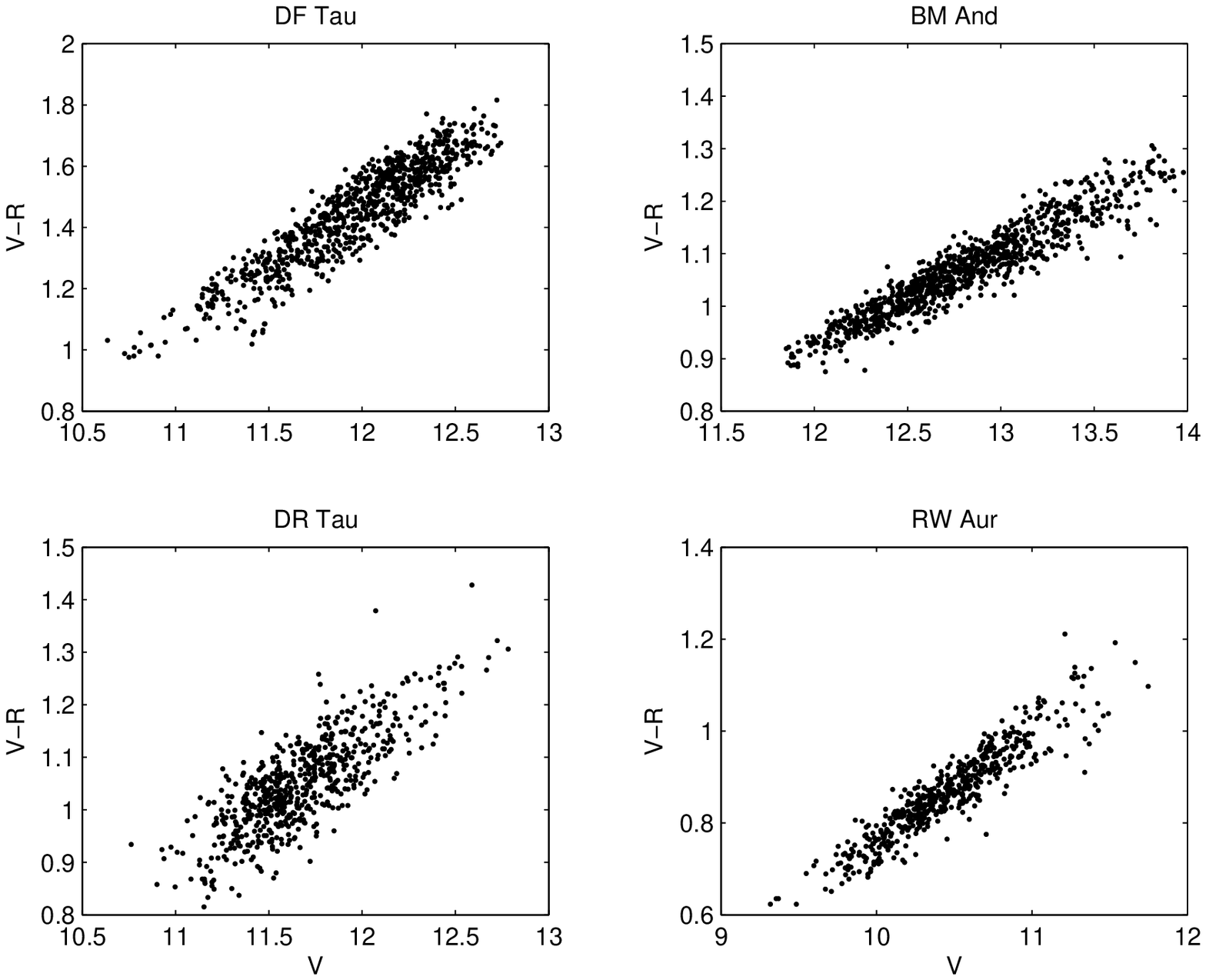} Fig.8. Illustration of the
  linear relation found for most objects between the $V-R$ color and $V$
  magnitude. Color slopes and correlation factors are listed in Table 2.
\end{minipage}
\end{figure*}

\begin{figure*}
\begin{minipage}[t]{8.7cm}
\includegraphics[width=8.5cm]{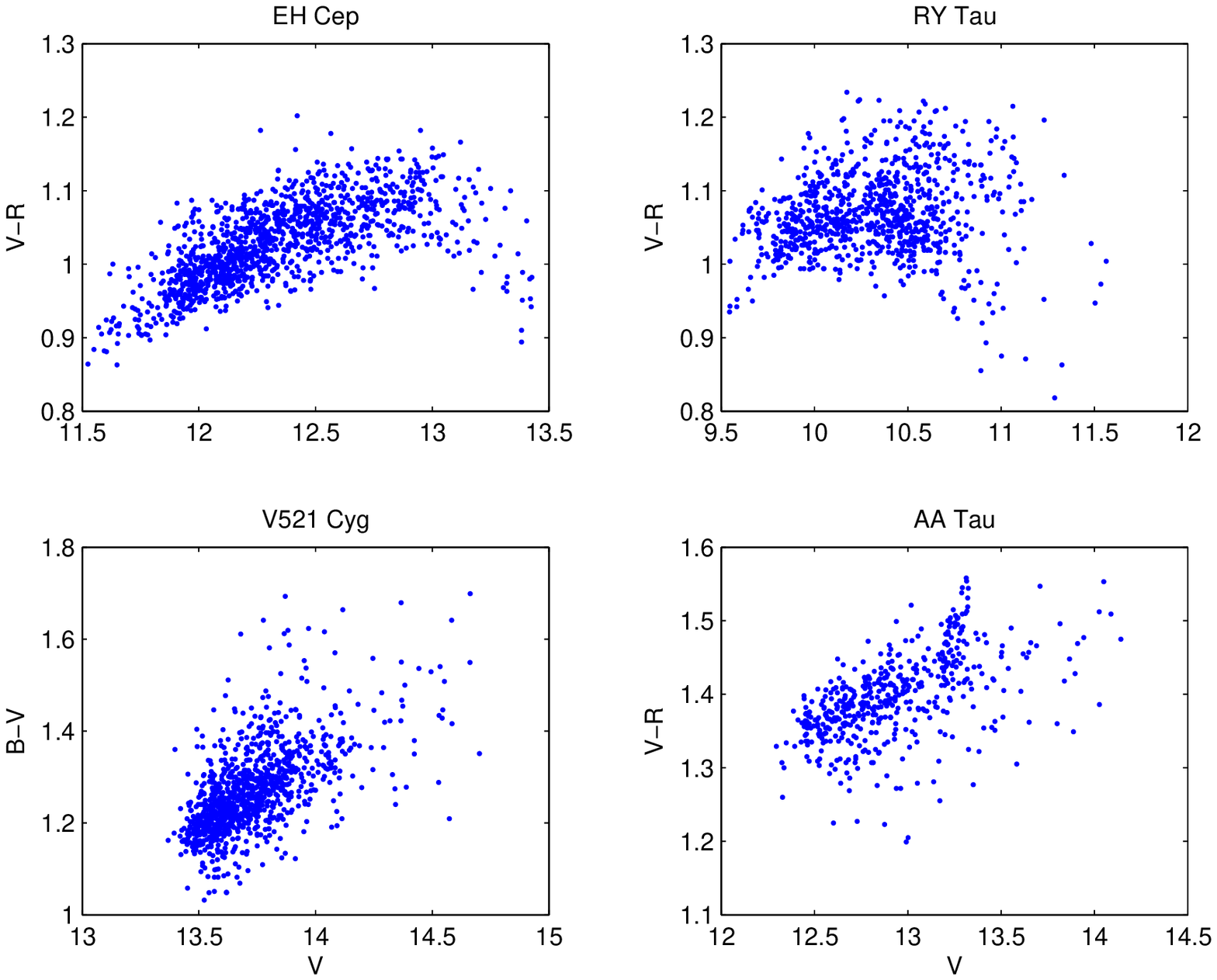}
Fig.9. Illustration of the non linear $V-R$ color variations on $V$
magnitude found for a few objects.
\end{minipage}
\hfill
\begin{minipage}[t]{8.7cm}
\includegraphics[width=8.5cm]{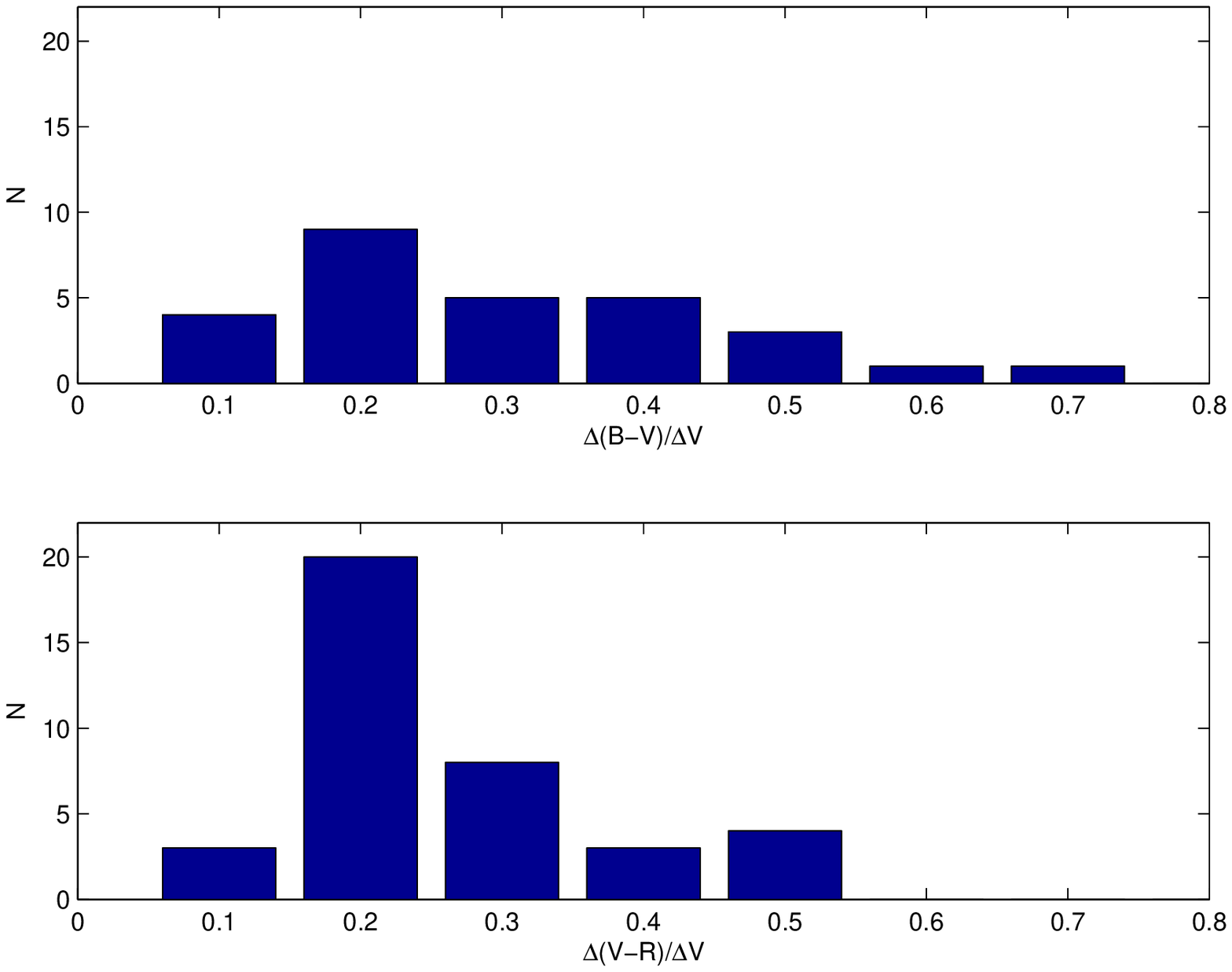}
Fig.10. Least-square color slopes. Top panel: $\frac{\Delta (B-V)}{\Delta
V}$; lower panel: $\frac{\Delta (V-R)}{\Delta V}$.
\end{minipage}
\end{figure*}

We define 2 additional parameters in order to characterize the
preferred brightness state of each object and the long term variations
of its extreme (mimimum and maximum) light levels.  The first
parameter is defined as $C1 = <\frac{V_{med} - V_{min}}{\Delta V}>$
where $V_{med}$ is the median brightness of the object in a given
season, $V_{min}$ its maximum brightness level and $\Delta V$ its
photometric amplitude during this season. C1 is computed for each
season and then averaged over all seasons.  The C1 parameter thus
characterizes the preferred light level of the object on the long
term. Thus defined, low C1 values indicate objects spending more time
in the high luminosity state than in the low one. \citet{parenago54}
discussed a similar classification of irregular variables based on a
histogram of their measured brightnesses.

The second parameter is defined as $C2 =
\sigma_{V_{min}}/\sigma_{V_{max}}$ where $V_{min}$ is the maximum
brightness state and $V_{max}$ the minimum one for each season.  This
parameter thus characterizes the long term variability (years) of the
bright state relative to that of the faint state. The C2 parameter can
be useful to clarify the origin of the variability. For instance, in
CTTs where the variability primarily results from hot spots, the
minimum light level ($\simeq$ photospheric level) is expected to
remain nearly constant over the years, while the maximum light level
(star + hot spot) is expected to vary on the long term as the spot
properties (size and/or temperature) change. Hence, CTTs with hot
spots are expected to have large C2 values. Conversely, CTTs whose
variability is due to cold spots or circumstellar extinction will
exhibit a nearly stable maximum light level ($\simeq$ photospheric
flux) while the minimum light level is expected to vary as the cold
spot or the occulting blob size evolve over the years. Hence, CTTs
whose variability is due to cold spots and/or circumstellar extinction
are expected to have low C2 values.



C1 and C2 were computed for 35 CTTs observed over at least 7 seasons,
in order to derive a reliable estimate of
$\sigma_{V_{min}}/\sigma_{V_{max}}$. Figure~7 shows C1 against C2 and
reveals some correlated behavior of the 2 parameters. The group of
objects in the lower left part of the diagram spend most of their time
close to the bright state (low C1). They also exhibit a fairly
constant maximum light level, while large variations of minimum
brightness occur on the long term (see, e.g. \object{AA Tau} and
\object{GW Ori} in Fig.~2). This behavior is suggestive of non
uniform circumstellar extinction occurring on the short term (weeks)
with an amplitude which varies on the long term (years). In support of
this hypothesis, we note that 8 of 10 UX Ori objects monitored at Mt
Maidanak lie in the same region of the diagram (not shown). Also, the
low frequency of such objects (5/35) agrees well with the probability
of observing occultation events from the circumstellar disk as
computed by \citet{bertout00}. Alternatively, slowly evolving cold
surface spots may produce this pattern for highly inclined
objects. For this group of objects, the maximum light level
corresponds to the photospheric luminosity.



At the other extreme, objects in the upper right part of the diagram
spend most of their time close to the faint state (high C1) and their
maximum light level varies significantly on the long term (see,
e.g. \object{DF Tau} and \object{DN Tau} in Fig.~2). The variability
of these objects is probably dominated by hot surface spots whose
properties vary over the years. The photospheric luminosity of these
objects would thus be best measured close to minimum
brightness. Finally, most CTTs lie in the intermediate region of the
diagram. They do not have any preferred brightness state
(C1$\simeq$0.5) and their extreme brightness levels exhibit the same
amount of long term variability (C2$\simeq$1), and sometimes (but not
always) vary in a correlated fashion. The photometric variability of
these objects may result from a mixture of hot and cold surface spots
whose surface distribution and/or properties vary on the long term
(see, e.g. \object{DR Tau}, \object{T Tau}, \object{BM And},
\object{DI Cep} in Fig.~2). We show in an accompanying paper that
WTTS, whose photometric variability is due to slowly evolving cold
surface spots, lie in this intermediate region of the diagram as well.

Finally, we investigate the color behavior of the 49 CTTs in our
sample. For most objects, the colors (either $(V-R)$ or $(B-V)$)
scale linearly with brightness. Hence, the color slope can be computed
from a linear least square fit. For most CTTs (39), the correlation
coefficient between $V-R$ and $V$ variations is higher than $0.47$
(cf. Fig.8). Only 5 stars do not show any reliable dependence of $V-R$
on brightness (correlation factor of lower than $0.47$) when averaged
over all seasons. These stars are: \object{GG Tau},
\object{LkH$\alpha$ 191}, \object{LkH$\alpha$ 321}, \object{LX Ori},
and \object{V360 Ori}. In addition, 5 CTTs demonstrate a more complex
color behavior, some exhibiting a blue turnaround towards minimum
brightness (cf. Fig.9). The unique color pattern of these stars
(\object{V521 Cyg}, \object{RY Tau}, \object{EH Cep}, \object{AA Tau},
and \object{CO Ori}) is discussed below.

The histogram of $\frac{\Delta{(V-R)}}{\Delta V}$ color slopes for
the 39 CTTs with linear color variations is shown in Figure~10. The
histogram peaks at values of order of 0.2--0.3, close to the value
expected for variable circumstellar extinction (0.25), assuming
interstellar extinction properties. This could be taken as
evidence that CTTs light variability is caused by a changing
screen of circumstellar dust intercepting the line of sight to the
star \citep[e.g.][]{bertout00}. However, similar color slopes can
be reproduced as well by other mechanisms. We show in an
accompanying paper that WTTS exhibit the same average
$\frac{\Delta{(V-R)}}{\Delta V}$ color slope as CTTs. In WTTS,
brightness and color variations result from the rotational
modulation of the stellar flux by cold surface spots. Hence, color
slopes usually do not uniquely distinguish between several
possible sources of variability.

A similar analysis was carried out for the
$\frac{\Delta{(B-V)}}{\Delta V}$ color slope, whose histogram is
also shown in Figure~10 and looks quite similar to the redder
color, with a maximum at 0.2. The ISM extinction law for
$\frac{\Delta{(B-V)}}{\Delta V}$ predicts a value of 0.32 if
color changes are due to variable circumstellar extinction.
However, besides extinction, other sources may affect the bluer
colors of CTTs, for example the well known veiling phenomenon
which is most important in the blue. Here again, color slopes are
not sufficient to distinguish between the various possible sources
of photometric variations.

\begin{figure*}
  \includegraphics[width=17.0cm]{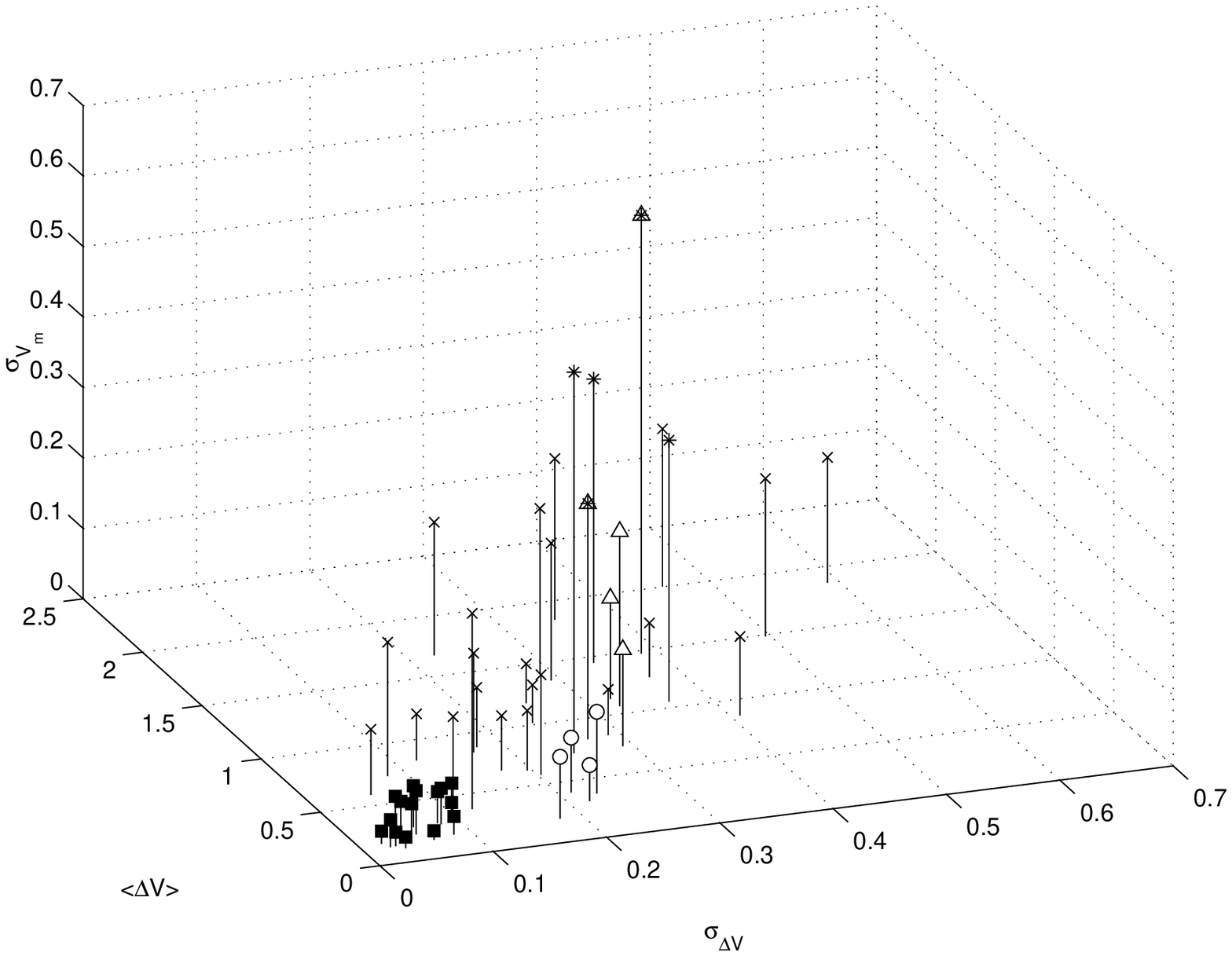} Fig.11. A 3D plot of the
  statistical properties of the long term light variability of 49 CTTs.
  The compact group of stars located close to the origin (filled squares)
  exhibit the lowest degree of variability. Five stars demonstrate large
  changes in their mean brightness level over the years (asterisks).
  Another small group exhibit significant seasonal variations in their
  photometric amplitude (circles). Five CTTs exhibiting unusual color
  behavior with a blue turnaround at minimum brightness are shown as
  triangles.

\end{figure*}

\section{Discussion}

The short term (weeks) photometric variability of CTTs is usually
interpreted as resulting from 3 possible causes, sometimes present
simultaneously: (i) cold surface spots, presumably resulting from
strong photospheric magnetic fields; (ii) hot spots at the stellar
surface, related to disk accretion onto the star; and (iii) variable
obscuration of the star by circumstellar matter.

The results obtained from the statistical analysis of the long
term light curves of 49 CTTs are graphically summarized in the
3D plot shown in Fig.~11. From these results, we attempt to
identify sub-groups of stars which appear to share the same type
of long term photometric behavior and try to relate it to
specific physical processes. A major difficulty is that
different processes may be at work on different timescales
(hours, days, weeks, years), and several sources of variability
are likely to coexist. In addition, other parameters, such as
the system inclination, may yield a different photometric
pattern for a given source of variability.  We thus merely try
to identify the dominant source of variability for each
subgroup, among the 49 stars with sufficient data.

The largest group of CTTs we identify in Fig.~11 consists of 37 stars which
exhibit homogeneous long term photometric variability. Of these, 15 are
located close to the origin of the plot exhibiting a low level of
variability. These 15 stars are shown by filled squares in Fig.~11 and the
other 22 stars as crosses. The spectral types of these stars span a range
from G5 up to M0. This group includes, e.g., \object{T Tau} and \object{DI
  Cep}.  CTTs in this group are characterized by relatively stable
patterns of variability on the long term. Many are known to be actively
accreting from their disk and the analysis of their short term (weeks)
photometric variability suggests it is dominated by hot surface spots,
though cold spots as well as circumstellar extinction can also be present
\citep{bouvier89, vrba93, bouvier95, bertout00}. Hence, the long term
variability of these stars merely reflects variability on shorter
timescales, with no significant variations from season to season.

The constancy of the light curves of these objects over the years
suggests that even though the properties of surface hot spots appear
to vary on the short term, possibly due to the changing accretion rate
onto the star, the accretion process remains active and fairly stable
on a timescale of several years. The large range of photometric
amplitude displayed by CTTs in this group, from a $0^m.1$ up to $1^m6$
in the V-band, may reflect either different accretion rates onto the
star or geometric projection effects resulting from the combination of
the spots latitude and the system's inclination. We do not find any
correlation between photometric amplitudes and the time-averaged
accretion rate in this sample using the H$_\alpha$ flux as a proxy, which
favors the latter explanation.

Besides this main group, a much smaller subgroup of deviant CTTs
seen in Fig.~11 exhibits large changes in their mean brightness
level over the years (\object{DG Tau}, \object{HN Tau},
\object{UY Aur}, \object{CO~Ori}, and \object{RY Tau}). These
stars are shown by asterisks in Fig.~11. Some of these, like
\object{UY Aur}, display a smooth change of mean brightness over
the years (Fig.~1). Others however, exhibit abrupt mean
brightness variations from one season to the next (e.g. from
$11^m.71$ to $13^m.31$ between 1997 and 1998 for \object{DG
Tau}). The origin of these variations cannot be unambiguously
assigned to a single mechanism. Large scale variations of the
accretion rate onto the star could produce such a behavior.
Alternatively, a variation of the circumstellar extinction on
the line of sight on a timescale of years cannot be excluded.
Such a timescale would then point to an obscuring screen located
in the outer disk regions.

Another small group exhibits significant seasonal variations in
their photometric amplitude (\object{EZ Ori}, \object{SU Aur},
\object{V2062 Oph}, and \object{GW Ori}). We have shown 
these stars as circles in Fig.~11. The long term light curve of these
objects is characterized by a nearly constant maximum brightness
level with a usually small amplitude of variability, but
interrupted at times by deep fading episodes. The amplitude
changes from season to season can amount to 52--90\% of the
average amplitude in these objects, while it amounts to only
10--47\% in the rest of the CTTs sample. Some of these objects
might be eclipsing binaries (e.g. \object{GW Ori}). Others may
undergo partial occultation by circumstellar disk material on a
timescale ranging from weeks to years (e.g. \object{SU Aur}, cf.
\citealt{dewarf03}, \citealt{bertout00}).

Finally, a last group of a few CTTs exhibit unusual color behavior with a
blue turnaround at minimum brightness (\object{V521 Cyg}, \object{RY Tau},
\object{EH Cep}, \object{AA Tau}, and \object{CO~Ori}). These stars are
shown as triangles in Fig.~11. This likely points to variable circumstellar
extinction~: as the stellar photosphere is partly occulted by circumstellar
material, the ratio of scattered light to direct light increases, and the
system turns bluer. These stars tend to be among the earlier type stars in
our sample, with spectral types from F8 to K2, except for AA Tau, quite a
typical CTTs with a spectral type K7 (\citealt{bouvier99, bouvier03, bouvier06}). Such
occultation effects are expected to be observed in only a small fraction of
CTTs seen at high inclination (\citealt{bertout00}).  Two stars of this
group (\object{RY Tau} and \object{CO~Ori}) also belong to the small group
of CTTs with large changes in their mean brightness level over the years.
They are shown by two superimposed symbols (asterisks and triangles) in
Fig.~11.

Our 20-year long photometric database thus provides the first
opportunity to investigate the long term variability of a significant
sample of CTTs. Yet, this baseline remains short compared to the
accretion (viscous) timescale in an accretion disk, not to mention the
Kelvin-Helmoltz PMS evolution timescale. Hence, while the longest
photometric database available today on PMS stars, it may still miss
some sources of variability acting on even longer timescales. As an
illustration, the 40-yr long light curve of T Tau investigated by
\citet{melnikov05} exhibits a gradual rise of the mean light level from
V$\simeq$10.5 to V$\simeq$9.9 over a timescale of about 20 years prior
to 1980.

\section{Conclusion}

The photometric variations of a large sample of CTTs have been
monitored at Mt. Maidanak Observatory. More than 21\,000 $UBVR$
photometric measurements have been collected in a uniform manner
for 72 CTTs over as many as twenty years. Based on this unique
database, we investigated here the long term photometric
properties of a subsample of 49 CTTs that were observed at least
12 times in each of 5 or more seasons. By defining a set of
statistical parameters, we characterized the long term photometric
patterns of this sample of CTTs.

Several types of long term photometric patterns were identified.  The
most common type, shared by about 75\% of the CTTs in our sample, is
characterized by low to moderate amplitudes of variability and
relatively stable light variations on a timescale of several years. We
argue that the source of variability in these objects is likely
dominated by a mixture of hot and cold surface spots. In this
subgroup, 15 stars were previously investigated by \citet{herbst94}
and classified as Type II variables. Even though the properties of the
spots may change on the short term (weeks), the overall stability of
the light variations over the years suggest that the underlying
physical processes, i.e., accretion onto the star for hot spots and
strong magnetic fields for cold spots, remain instrumental and undergo
only marginal changes on the longer term.

In addition to this common pattern, a few more peculiar types of long
term photometric variations are found. They are characterized by
significant changes in the mean brightness level and/or photometric
amplitudes over the years. We suggest that, in most cases, this type
of unusual variability is dominated by variable circumstellar
extinction, presumably due to the inner (short term) or outer (long
term) disk regions partially occulting the central star. Occultation
effects are likely to happen only in objects seen at high inclination,
which explains why these patterns are more rarely observed
(cf. \citet{bertout00}), being exhibited by only 12 CTTs over 49 in
our sample. Most of the stars in this subgroup correspond to Type III
variability as defined by \citet{herbst94}. As previously noted by
these authors, we find that most (8/12) of the CTTs exhibiting this
peculiar type of variability have a spectral type K2 or earlier, while
most stars in our sample have later spectral types. Hence, this type
of variability seems to preferentially (but not exclusively) affect
early-type CTTs and may be related to the UXor variability pattern
seen in a minority of the more massive Herbig Ae-Be stars.

Finally, we caution that photometric variability in CTTs, be it short
or long term, often results from several processes acting
simultaneously and which may be difficult to identify from photometry
alone. Spectroscopic and polarimetric measurements obtained
simultaneously with photometric ones, which have been gathered so far
for very few CTTs, usually offer a clearer view of the physical
processes responsible for the observed variability.

The results reported here on the long term photometric behavior of CTTs
will be contrasted with a similar analysis of the long term photometric
variations of WTTs from the same database in an accompanying paper.


\begin{acknowledgements}
The authors wish to thank M. Ibrahimov, S. Yakubov, O. Ezhkova, V.
Kondratiev and many other observers and students, who worked or are
still working in the Tashkent Astronomical Institute, for their
participation in the photometric observations. We are particularly
grateful to C. Dougados, C. Bertout, and F. M\'enard for their
comments and suggestions pertaining to this work. We thank an
anonymous referee for comments that helped to improve the manuscript.
Support of the American Astronomical Society, European Southern
Observatory (grant A-02-048), and International Science Foundation by
Soros (grant MZA000) and a CRDF grant (ZP1-341) is acknowledged. This
work was also supported by a NATO Collaborative Linkage grant for
European countries (PST.CLG.976194) and by an ECO-NET program of the
French Ministry of Foreign Affairs, which we both gratefully
acknowledge. W.H. gratefully acknowledges the support of NASA through
its Origins of Solar Systems Program.
\end{acknowledgements}

\begin{longtable}{lrllrlrll}
      \caption[]{Results of Maidanak long-term photometry of CTTs}\\
\hline\hline
     Name  &   HBC  &        SpT & JD$_{min}$--JD$_{max}$ &         $N_s$ &    $V$ range &       $N_{obs}$ &      $\overline{B-V}$ &      $\overline{V-R}$ \\
\hline
\endfirsthead
\caption{continued.}\\
\hline\hline
     Name  &   HBC  &        SpT & JD$_{min}$--JD$_{max}$ &         $N_s$ &    $V$ range &       $N_{obs}$ &      $\overline{B-V}$ &      $\overline{V-R}$ \\
\hline
\endhead
\hline                  
    AA Ori &        130 &         K4 & 47055-47777 &          1 & 12.79-13.16 &         11 &       1.28 &       1.23 \\
    AA Tau &         63 &         K7 & 46690-53028 &         15 & 12.30-14.14 &        568 &       1.34 &       1.40 \\
    AS 205 &        254 &         K5 & 46955-49588 &          8 & 10.87-13.77 &        362 &       1.24 &       1.55 \\
    AV Ori &        159 &         K4 & 47040-47173 &          1 & 12.84-14.42 &         14 &       1.23 &       1.23 \\
BD-22 4059 &        608 &         G5 & 47670-49514 &          5 & 10.10-10.28 &
165 &       0.84 &       0.75 \\
    BM And &        318 &         K5 & 47008-52908 &         17 & 11.85-13.98 &        999 &       1.13 &       1.07 \\
    BP Tau &         32 &         K7 & 46679-52938 &         11 & 11.67-12.99 &        495 &       1.06 &       1.26 \\
    CI Tau &         61 &         K7 & 47018-52935 &          7 & 12.28-13.79 &        319 &       1.37 &       1.55 \\
    CO Ori &         84 &         F8 & 47031-52927 &         16 & 9.81-12.73 &        541 &       1.06 &       0.99 \\
    CX Tau &         27 &       M2.5 & 51782-51841 &          1 & 13.62-13.80 &         15 &       1.51 &       1.65 \\
    CY Tau &         28 &         M1 & 52579-53028 &          1 & 12.94-13.59 &         19 &       1.07 &       1.53 \\
    DD Tau &         30 &         M1 & 48123-49275 &          4 & 13.45-15.15 &        114 &       1.07 &       1.76 \\
    DE Tau &         33 &         M2 & 46681-47382 &          1 & 12.82-13.95 &         23 &       1.43 &       1.64 \\
    DF Tau &         36 &         M0 & 45920-53028 &         19 & 10.64-12.74 &        929 &       1.10 &       1.45 \\
    DG Tau &         37 &          M & 45928-52938 &         18 & 11.23-13.58 &        710 &       1.06 &       1.35 \\
    DH Tau &         38 &        M1e & 52932-52938 &          1 & 13.20-13.73 &          5 &       0.89 &       1.36 \\
    DI Cep &        315 &         G8 & 46966-52907 &         17 & 11.14-11.67 &       1231 &       0.89 &       0.88 \\
    DI Tau &         39 &         M0 & 46710-51841 &         11 & 12.69-12.97 &        435 &       1.59 &       1.50 \\
    DK Tau &         45 &         K7 & 47022-52937 &         11 & 11.54-14.46 &        429 &       1.27 &       1.43 \\
    DL Tau &         58 &         K7 & 47017-52935 &          8 & 12.39-13.89 &        370 &       1.09 &       1.42 \\
    DM Tau &         62 &         M1 & 47028-52935 &          6 & 13.43-14.32 &        162 &       1.07 &       1.51 \\
    DN Tau &         65 &         M0 & 46681-52914 &         10 & 12.08-12.54 &        369 &       1.33 &       1.35 \\
    DR Tau &         74 &         K5 & 46686-52924 &         16 & 10.76-12.78 &        642 &       0.82 &       1.05 \\
    DS Tau &         75 &         K5 & 46680-48230 &          4 & 11.58-12.68 &        199 &       0.91 &       1.09 \\
    EH Cep &        307 &         K2 & 46970-52907 &         17 & 11.52-13.44 &       1296 &       1.19 &       1.03 \\
    EZ Ori &        114 &         G0 & 47048-53022 &          7 & 11.48-12.25 &        222 &       0.84 &       0.77 \\
    GG Tau &         54 &         K7 & 46686-49287 &          7 & 11.91-12.48 &        345 &       1.42 &       1.44 \\
    GI Tau &         56 &         K6 & 46681-52935 &          8 & 12.30-14.85 &        344 &       1.46 &       1.57 \\
    GK Tau &         57 &         K7 & 46682-52935 &          8 & 11.94-13.58 &        376 &       1.43 &       1.44 \\
    GM Aur &         77 &         K3 & 46681-50046 &          9 & 11.74-12.35 &        349 &       1.21 &       1.22 \\
    GW Ori &         85 &         G5 & 47031-52927 &         15 & 9.74-10.53 &        531 &       0.99 &       0.94 \\
    GX Ori &         89 &         K1 & 47031-47169 &          1 & 13.11-13.47 &         20 &       0.95 &       1.02 \\
  Haro 1-1 &        256 &       K5.7 & 46964-49190 &          6 & 12.80-13.82 &        166 &       1.27 &       1.45 \\
 Haro 1-14 &        267 &         M0 & 46964-50660 &          6 & 13.75-15.03 &        154 &       1.52 &       1.68 \\
  Haro 1-4 &        257 &       K6.7 & 46964-48837 &          5 & 13.29-13.81 &        131 &       1.75 &       1.85 \\
  Haro 1-8 &        261 &         K5 & 46965-49190 &          4 & 13.74-14.22 &        133 &       1.69 &       1.70 \\
 Haro1-14c &        644 &         K3 & 46607-46649 &          1 & 12.13-13.13 &         27 &       1.67 &            \\
    HI Ori &         93 &         K4 & 47030-52176 &          7 & 12.63-13.86 &        214 &       1.28 &       1.20 \\
    HL Tau &         49 &     K7-M2? & 48147-48954 &          2 & 14.22-14.63 &         48 &       1.42 &       1.54 \\
    HN Tau &         60 &         K5 & 47028-49277 &          6 & 12.63-14.94 &        163 &       0.97 &       1.30 \\
    HP Tau &         66 &         K3 & 46702-47421 &          1 & 13.35-14.20 &         35 &       1.78 &       1.69 \\
   LkCa  8 &        385 &         M0 & 48953-52935 &          1 & 12.83-13.44 &         26 &       1.31 &       1.39 \\
   LkCa 15 &        419 &         K5 & 48859-50760 &          5 & 11.85-13.34 &        123 &       1.26 &       1.16 \\
  LkHa 172 &        298 &         K4 & 48087-49962 &          2 & 14.58-14.98 &         64 &       1.44 &       1.36 \\
  LkHa 191 &        301 &         K0 & 46618-49956 &          5 & 12.87-13.09 &        326 &       1.10 &       1.04 \\
  LkHa 200 &          6 &         K1 & 46258-46802 &          1 & 13.51-13.68 &         29 &       1.23 &       1.10 \\
  LkHa 209 &        194 &         G8 & 46722-50044 &          8 & 11.64-12.06 &        193 &       0.83 &       0.80 \\
  LkHa 228 &        295 &       K1.2 & 46261-50017 &          2 & 12.87-13.06 &         48 &       0.84 &       0.78 \\
  LkHa 274 &        203 &         K4 & 46722-50024 &          3 & 13.16-13.49 &         92 &       1.26 &       1.20 \\
  LkHa 321 &        303 &          G & 46985-50020 &          7 & 12.25-12.44 &        547 &       1.23 &       1.17 \\
    LX Ori &        133 &         K3 & 47039-51834 &          6 & 11.98-12.34 &        191 &       1.10 &       1.00 \\
     Oph 6 &        653 &          K & 48419-49245 &          1 & 13.18-13.45 &         48 &       1.69 &       1.75 \\
  RW AurAB &      80.81 &      K1-K3 & 46680-52924 &         16 & 9.32-11.75 &        534 &       0.72 &       0.87 \\
    RY Tau &         34 &         K1 & 45968-53022 &         19 & 9.55-11.56 &        885 &       1.03 &       1.07 \\
     S CrA &        286 &         K6 & 47673-48838 &          3 & 10.40-12.32 &        122 &       1.00 &       1.18 \\
    SU Aur &         79 &         G2 & 45676-52926 &         19 & 8.92-10.02 &        793 &       0.92 &       0.83 \\
    SW Ori &        115 &         K4 & 47041-50022 &          8 & 12.42-13.69 &        153 &       1.04 &       0.97 \\
     T Tau &         35 &         K0 & 46665-52937 &         17 & 9.75-10.18 &        982 &       1.17 &       1.13 \\
  UX TauAB &      42.43 &      M1-K2 & 47378-49287 &          6 & 10.64-12.77 &        318 &       1.17 &       1.06 \\
    UY Aur &         76 &         K7 & 46681-52914 &         14 & 11.28-13.68 &        427 &       1.27 &       1.43 \\
    UZ Tau &         52 &       M1.3 & 46383-47101 &          1 & 12.25-13.11 &         15 &       1.19 &       1.66 \\
 V1082 Cyg &        728 &            & 48059-50022 &          6 & 12.62-13.67 &        461 &       0.76 &       1.00 \\
 V1121 Oph &        270 &         K5 & 46955-52876 &         14 & 10.80-11.83 &        601 &       1.28 &       1.36 \\
 V2058 Oph &        259 &         K6 & 46607-49194 &          7 & 12.60-13.09 &        209 &       1.49 &       1.60 \\
 V2062 Oph &        268 &         K2 & 46607-49194 &          7 & 12.20-13.05 &        223 &       1.34 &       1.39 \\
  V360 Ori &        144 &         K6 & 47060-50022 &          5 & 12.40-12.71 &        159 &       1.39 &       1.30 \\
  V521 Cyg &        299 &         G8 & 46609-52899 &         17 & 13.37-14.70 &       1140 &       1.27 &       1.20 \\
  V625 Ori &        183 &         K6 & 47061-50010 &          5 & 12.61-13.97 &        114 &       1.08 &       1.15 \\
  V649 Ori &         86 &         G8 & 47031-52176 &          7 & 11.87-12.31 &        275 &       1.12 &       1.09 \\
  V853 Oph &        266 &       M1.5 & 46607-51014 &          7 & 12.61-13.87 &        238 &       1.10 &       1.63 \\
    XZ Tau &         50 &         M3 & 48142-48954 &          3 & 12.96-15.17 &         74 &       1.40 &       1.88 \\
    YY Ori &        119 &      K5.M0 & 47041-47170 &          1 & 13.20-14.08 &         16 &       0.77 &       1.01 \\
\hline
\end{longtable}

\begin{table*}
\caption[]{Statistical properties of CTTs light curves (see text).}\centering
\begin{tabular}{lrrllllllllll}
\hline Star Name &    $N_s^*$    &$\overline{V_m}$& $\sigma_{V_m}$
&$\overline{\Delta V}$ & $\sigma_{\Delta V}$ &
$\frac{\sigma_{\Delta V}}{\overline{\Delta V}}$ &
$\frac{\Delta(B-V)}{\Delta V}$ & $\rho_{B-V}$ &
$\frac{\Delta(V-R)}{\Delta V}$ & $\rho_{V-R}$ & $ c1 $ & $ c2 $\\
\hline
    AA Tau & 15 & 12.917 & 0.143 & 1.133 & 0.322 & 0.285 & 0.017 &  0.04 & 0.093 &  0.54 & 0.423 &  0.304\\
    AS 205 &  8 & 12.259 & 0.224 & 2.002 & 0.459 & 0.229 & 0.129 &  0.51 & 0.239 &  0.89 & 0.997 &  0.452\\
 BD-224059 &  5 & 10.169 & 0.016 & 0.116 & 0.035 & 0.300 & 0.073 &  0.08 & 0.483 &  0.53 & 0.525 &  0.427\\
    BM And & 17 & 12.809 & 0.189 & 1.673 & 0.223 & 0.133 & 0.135 &  0.82 & 0.177 &  0.94 & 0.740 &  0.442\\
    BP Tau & 11 & 12.130 & 0.085 & 0.630 & 0.196 & 0.311 & 0.531 &  0.68 & 0.363 &  0.84 & 0.450 &  0.499\\
    CI Tau &  7 & 13.107 & 0.085 & 0.874 & 0.177 & 0.203 & 0.340 &  0.72 & 0.300 &  0.85 & 1.635 &  0.532\\
    CO Ori & 15 & 10.960 & 0.622 & 1.473 & 0.385 & 0.261 & 0.061 &  0.56 & 0.105 &  0.84 & 0.668 &  0.425\\
    DF Tau & 19 & 11.983 & 0.195 & 1.345 & 0.292 & 0.217 & 0.392 &  0.83 & 0.386 &  0.92 & 1.642 &  0.629\\
    DG Tau & 18 & 12.207 & 0.541 & 0.723 & 0.247 & 0.341 & 0.144 &  0.74 & 0.187 &  0.91 & 1.032 &  0.484\\
    DI Cep & 17 & 11.389 & 0.059 & 0.280 & 0.059 & 0.211 & 0.190 &  0.46 & 0.178 &  0.48 & 0.973 &  0.561\\
    DI Tau &  9 & 12.843 & 0.013 & 0.156 & 0.064 & 0.409 & 0.147 &  0.09 & 0.440 &  0.57 & 0.998 &  0.428\\
    DK Tau & 11 & 12.583 & 0.178 & 1.863 & 0.590 & 0.316 & 0.151 &  0.58 & 0.223 &  0.90 & 0.250 &  0.444\\
    DL Tau &  9 & 13.146 & 0.054 & 1.015 & 0.241 & 0.238 & 0.189 &  0.61 & 0.300 &  0.85 & 1.176 &  0.531\\
    DM Tau &  6 & 13.928 & 0.093 & 0.591 & 0.054 & 0.091 & 0.596 &  0.79 & 0.472 &  0.83 & 1.285 &  0.571\\
    DN Tau & 10 & 12.332 & 0.045 & 0.289 & 0.081 & 0.279 & 0.447 &  0.46 & 0.223 &  0.54 & 2.177 &  0.542\\
    DR Tau & 16 & 11.660 & 0.077 & 1.271 & 0.371 & 0.292 & 0.173 &  0.66 & 0.229 &  0.80 & 0.617 &  0.432\\
    EH Cep & 17 & 12.391 & 0.248 & 1.064 & 0.323 & 0.303 & 0.046 &  0.38 & 0.095 &  0.63 & 0.987 &  0.423\\
    EZ Ori &  7 & 11.677 & 0.078 & 0.404 & 0.211 & 0.522 & 0.256 &  0.79 & 0.236 &  0.85 & 0.369 &  0.350\\
    GG Tau &  7 & 12.212 & 0.039 & 0.328 & 0.048 & 0.145 & 0.696 &  0.50 & 0.139 &  0.30 & 1.375 &  0.538\\
    GI Tau &  8 & 13.392 & 0.229 & 1.840 & 0.347 & 0.189 & 0.231 &  0.58 & 0.269 &  0.92 & 1.014 &  0.377\\
    GK Tau &  8 & 12.668 & 0.056 & 1.184 & 0.253 & 0.214 & 0.180 &  0.42 & 0.226 &  0.82 & 0.747 &  0.382\\
    GM Aur &  9 & 12.052 & 0.039 & 0.378 & 0.103 & 0.273 & 0.080 &  0.12 & 0.369 &  0.71 & 1.059 &  0.500\\
    GW Ori & 15 &  9.878 & 0.088 & 0.200 & 0.180 & 0.901 & 0.159 &  0.54 & 0.210 &  0.79 & 0.171 &  0.360\\
  Haro 1-1 &  6 & 13.355 & 0.078 & 0.656 & 0.176 & 0.268 & 0.390 &  0.40 & 0.214 &  0.52 & 1.293 &  0.520\\
 Haro 1-14 &  6 & 14.320 & 0.141 & 0.833 & 0.170 & 0.203 & 0.072 &  0.09 & 0.156 &  0.48 & 1.086 &  0.433\\
  Haro 1-4 &  5 & 13.529 & 0.018 & 0.291 & 0.107 & 0.367 & 0.366 &  0.19 & 0.464 &  0.59 & 0.478 &  0.609\\
    HI Ori &  7 & 13.347 & 0.278 & 0.368 & 0.120 & 0.326 & 0.135 &  0.31 & 0.195 &  0.70 & 0.930 &  0.469\\
    HN Tau &  6 & 13.993 & 0.403 & 1.446 & 0.340 & 0.235 & 0.055 &  0.33 & 0.110 &  0.66 & 1.069 &  0.612\\
   LkCa 15 &  5 & 12.144 & 0.065 & 0.837 & 0.289 & 0.346 & 0.250 &  0.88 & 0.188 &  0.89 & 0.208 &  0.263\\
  LkHa 191 &  5 & 12.960 & 0.018 & 0.178 & 0.020 & 0.115 & 0.211 &  0.21 & 0.063 &  0.06 & 0.841 &  0.421\\
  LkHa 209 &  8 & 11.858 & 0.026 & 0.179 & 0.084 & 0.469 & 0.241 &  0.31 & 0.313 &  0.49 & 1.022 &  0.435\\
  LkHa 321 &  7 & 12.342 & 0.020 & 0.143 & 0.029 & 0.200 & 0.034 &  0.05 & 0.118 &  0.22 & 0.599 &  0.492\\
    LX Ori &  6 & 12.091 & 0.039 & 0.140 & 0.024 & 0.170 & 0.031 &  0.06 & 0.182 &  0.40 & 0.800 &  0.466\\
  RW AurAB & 16 & 10.516 & 0.235 & 1.411 & 0.289 & 0.205 & 0.189 &  0.74 & 0.229 &  0.93 & 0.941 &  0.488\\
    RY Tau & 11 & 10.346 & 0.334 & 0.827 & 0.270 & 0.326 & 0.033 &  0.21 & 0.018 &  0.11 & 0.956 &  0.462\\
    SU Aur & 19 &  9.239 & 0.116 & 0.368 & 0.230 & 0.624 & 0.269 &  0.88 & 0.240 &  0.88 & 0.414 &  0.354\\
    SW Ori &  5 & 12.980 & 0.190 & 0.728 & 0.083 & 0.114 & 0.181 &  0.74 & 0.202 &  0.84 & 1.107 &  0.326\\
     T Tau & 17 &  9.972 & 0.062 & 0.220 & 0.055 & 0.252 & 0.240 &  0.50 & 0.203 &  0.54 & 1.191 &  0.541\\
  UX TauAB &  6 & 11.286 & 0.224 & 1.487 & 0.496 & 0.333 & 0.251 &  0.96 & 0.246 &  0.98 & 0.138 &  0.342\\
    UY Aur & 13 & 12.560 & 0.371 & 1.050 & 0.365 & 0.348 & 0.385 &  0.86 & 0.314 &  0.89 & 0.973 &  0.544\\
 V1082 Cyg &  6 & 13.142 & 0.066 & 0.828 & 0.119 & 0.144 & 0.356 &  0.84 & 0.222 &  0.78 & 0.578 &  0.552\\
 V1121 Oph & 14 & 11.430 & 0.132 & 0.385 & 0.105 & 0.272 & 0.427 &  0.72 & 0.164 &  0.62 & 1.419 &  0.569\\
 V2058 Oph &  7 & 12.825 & 0.051 & 0.277 & 0.083 & 0.299 & 0.494 &  0.48 & 0.342 &  0.63 & 0.753 &  0.544\\
 V2062 Oph &  7 & 12.476 & 0.051 & 0.314 & 0.218 & 0.693 & 0.453 &  0.50 & 0.341 &  0.56 & 0.429 &  0.483\\
  V360 Ori &  5 & 12.529 & 0.056 & 0.152 & 0.044 & 0.290 & 0.311 &  0.37 & 0.149 &  0.39 & 0.525 &  0.496\\
  V521 Cyg & 17 & 13.736 & 0.137 & 0.732 & 0.291 & 0.397 & 0.275 &  0.61 & 0.246 &  0.71 & 0.291 &  0.289\\
  V625 Ori &  5 & 13.136 & 0.112 & 0.862 & 0.408 & 0.474 & 0.385 &  0.73 & 0.258 &  0.83 & 0.364 &  0.458\\
  V649 Ori &  7 & 12.078 & 0.042 & 0.262 & 0.046 & 0.176 & 0.249 &  0.41 & 0.238 &  0.62 & 0.671 &  0.504\\
  V853 Oph &  7 & 13.443 & 0.142 & 0.580 & 0.203 & 0.350 & 0.533 &  0.61 & 0.544 &  0.83 & 2.111 &  0.596\\
\hline\end{tabular}
\end{table*}



\bibliographystyle{aa}
\bibliography{paper}

\end{document}